%% file: root.tex
\documentclass[letterpaper, 10 pt, conference]{ieeeconf}  

\IEEEoverridecommandlockouts                              

\usepackage{graphicx} 
\usepackage{amsmath} 
\usepackage{microtype}
\usepackage[backgroundcolor=cyan]{todonotes}
\usepackage{units}
\usepackage{booktabs}
\usepackage{algorithm, algorithmic}
\usepackage{mathtools}
\usepackage[hidelinks]{hyperref}
\hypersetup{
    colorlinks=false,
    linkcolor=black,
    filecolor=black,      
    urlcolor=black
}
\usepackage[noadjust]{cite}

\usepackage{amssymb}
\usepackage{wrapfig}

\usepackage{tikz}
\def\checkmark{\tikz\fill[scale=0.35](0,.35) -- (.25,0) -- (1,.7) -- (.25,.15) -- cycle;}

\usepackage{tikz}
\usepackage{pgfplots}
\usetikzlibrary{arrows,decorations,backgrounds,shapes,chains}
\pgfplotsset{compat=newest}
\pgfplotsset{plot coordinates/math parser=false}
\newlength\figureheight
\newlength\figurewidth

\usepackage{tabularx}

\makeatletter
\newcommand\footnoteref[1]{\protected@xdef\@thefnmark{\ref{#1}}\@footnotemark}
\makeatother

\input{mathdef}

\title{\LARGE \bf
A Family of Iterative Gauss-Newton Shooting Methods for Nonlinear Optimal Control
}
\author{
Markus Giftthaler$^1$, 
Michael Neunert$^1$, 
Markus St\"auble$^1$,
Jonas Buchli$^1$
and Moritz Diehl$^2$ 
\thanks{
$^1$Agile \& Dexterous Robotics Lab, ETH Z\"urich, Switzerland. {\footnotesize \{mgiftthaler, neunertm, markusta, buchlij\}@ethz.ch} \newline
$^2$Systems Control and Optimization Laboratory, Department of Microsystems Engineering (IMTEK), University of Freiburg, Germany. {\footnotesize moritz.diehl@imtek.uni-freiburg.de }
}
}

\begin{document}
\maketitle
\thispagestyle{empty}
\pagestyle{empty}

\begin{abstract}
This paper introduces a family of iterative algorithms for unconstrained nonlinear optimal control. 
We generalize the well-known iLQR algorithm to different multiple-shooting variants, combining advantages like straight-forward initialization and a closed-loop forward integration. 
All algorithms have similar computational complexity, i.e. linear complexity in the time horizon, and can be derived in the same computational framework.
We compare the full-step variants of our algorithms and present several simulation examples, including a high-dimensional underactuated robot subject to contact switches.
Simulation results show that our multiple-shooting algorithms can achieve faster convergence, better local contraction rates and much shorter runtimes than classical iLQR, which makes them a superior choice for nonlinear model predictive control applications.
\end{abstract}
%
%
\section{Introduction}
\label{sec:Introduction}
\subsection{Overview and Motivation}
In this paper, we discuss a family of iterative Gauss-Newton shooting methods for numerically solving unconstrained optimal control problems, and illustrate the effectiveness of our algorithms with various robotics examples.
We outline the connection between a number of `direct' optimal control methods and Gauss-Newton methods from the class of Differential Dynamic Programming (DDP)~\cite{mayne1966ddp} algorithms.
Additionally, we present a natural extension arising from this connection and introduce a family of hybrid Gauss-Newton Multiple Shooting methods.

In direct approaches to optimal control, infinite-dimensional optimal control problems are transcribed into finite dimensional Nonlinear Programs (NLPs). Two prominent ways of transcription by `shooting' are direct single shooting (SS) and direct multiple shooting (MS)~\cite{bock1984direct}, which differ in the choice of decision variables.
In single shooting, solely the control inputs are the decision variables. Generally speaking, the control trajectory is discretized in a piece-wise polynomial fashion (for simplicity, we focus on piece-wise constant controls in this paper, c.f. Fig.~\ref{fig:algorithms}a). A corresponding state trajectory is obtained by means of numerical forward integration of the system dynamics, starting at a given initial state. SS is often called a `sequential' approach.
In multiple shooting, the same discretization scheme is employed for the control inputs, but additionally, intermediate states are added to the decision variables. This provides several advantages~\cite{diehl2006fast}, but requires the introduction of additional matching constraints to ensure continuity of the state trajectory. 
The technique of introducing these additional degrees of freedom into the original problem, combined with adding matching constraints, is called \emph{lifting}~\cite{albertsmeyerLifted}, and results in a `simultaneous' method.

The formulation of both SS and MS as standard NLPs is straightforward and any state-of-the-art NLP solvers can be used to solve them.
It is important to note that under the assumption of having a piece-wise polynomial control parameterization, the intrinsic sparsity structure of the underlying optimal control problem of both SS and MS allow them to achieve linear time complexity by performing a Riccati recursion~\cite{Rao1998application}.

Classical single shooting often does not perform well for unstable systems due to the pure open-loop forward integration of the system dynamics. In DDP, this is handled by doing a closed-loop forward integration, using a feedforward plus a time-varying state-feedback control law. The Riccati backward sweep designs time-varying feedback gains on the fly without additional computational cost. 
DDP is an exact-Hessian method, requiring the computation of second derivatives of the dynamics. While this gives the algorithm quadratic convergence, this can be impractical for use in systems with complex dynamics. For that reason, Hessian-approximating variants of DDP have become quite popular in the robotics community~\cite{koenemann2015whole,tassaSynthesis,neunert:2017:ral,ponton2016risk,farshidian16efficient}.

An important Hessian-approximating variant of DDP is the \emph{iterative Linear-Quadratic Regulator} (iLQR)~\cite{todorov2005ilqg}, which is also known as \emph{Sequential Linear Quadratic Optimal Control}~\cite{slq:2005}. This method can be classified as closed-loop single shooting using a Gauss-Newton Hessian approximation and a Riccati backward sweep to solve linear-quadratic (LQ) subproblems. The Gauss-Newton Hessian approximation is based on the assumption that the objective function can be locally approximated as a sum of quadratic terms, and requires only first-order derivatives of the system dynamics. This comes at the cost of giving only linear convergence, however.
The Gauss-Newton approach can be lifted, too, which has for example been shown in~\cite{schaeferThesis}. While it initially appears to be a drawback to increase the number of decision variables, it is important to emphasize that the lifted problem can be solved at approximately the same computational cost as the original non-lifted problem, and can lead to a significant increase of convergence speed~\cite{albertsmeyerLifted}.
Therefore, the fundamental motivation for this paper is to combine the benefits of iLQR with a multiple-shooting approach.

\subsection{Contribution}
\label{sec:contribution}
In this work, we derive a lifted equivalent of iLQR, called \emph{Gauss-Newton Multiple Shooting} (GNMS), which introduces the intermediate states as additional decision variables.
Next, we extend this relationship to form an entire family of open-loop multiple-shooting algorithms, denoted GNMS($M$), and closed-loop multiple shooting algorithms, denoted as iLQR-GNMS($M$). The latter is shown to be a generalization of iLQR and can be considered multiple-shooting iLQR. We outline the relationship between these algorithms and existing methods. We give simulation examples including a complex underactuated robot and compare the performance of the full-step algorithms using data gained from hardware experiments.
Furthermore, we show the benefits of iLQR-GNMS($M$) methods for nonlinear model predictive control.

\subsection{Outline}
This paper is structured as follows. 
In Section~\ref{sec:gnms}, we derive GNMS, and present the basic update routine for state and control trajectories. Using these update equations, we generalize iLQR and GNMS to a family of algorithms in Section~\ref{sec:algorithms}.
Section~\ref{sec:simulation_results} showcases several simulation results, based on data gained from hardware experiments.
A discussion and outlook concludes the paper in Section~\ref{sec:Discussion}.

\section{Gauss-Newton Multiple Shooting}
\label{sec:gnms}

\begin{figure}
\centering
\includegraphics[width=0.99\columnwidth]{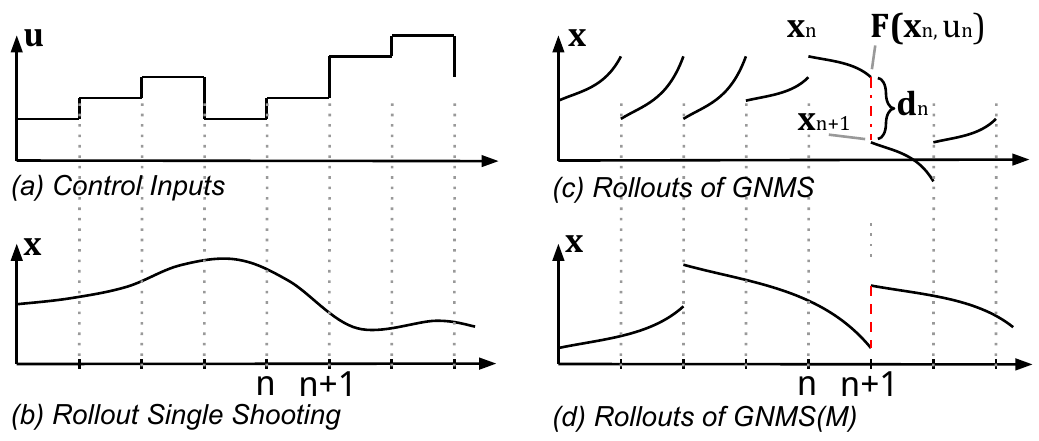}
\caption{Intuition about different shooting variants. 
(a) A zero-order hold control parameterization, with a constant control input at each stage.
(b) Single shooting, where the state trajectory is obtained through a single forward integration over the whole problem horizon. 
(c) In GNMS, intermediate states at every time-index $n$ are introduced as additional decision variables.
(d) In hybrid versions of GNMS, the multiple-shooting intervals span several control intervals. The intermediate states at the beginning of the multiple-shooting intervals are decision variables, and the states in between are obtained by forward integration.
}\label{fig:algorithms}
\end{figure}

In the following, a brief derivation of the unconstrained Gauss-Newton Multiple Shooting method is presented. We show the derivation using an intuitive value-function approach in the style of~\cite{todorov2005ilqg} in order to highlight the close relationship between GNMS and iLQR. However, from the beginning, we lift the optimization problem and introduce intermediate states as additional decision variables besides the controls. In that sense, having each control decision variable accompanied with a state-decision variable, GNMS is closely related to the original multiple-shooting algorithm~\cite{bock1984direct}.

Consider the following discrete-time, finite-horizon, nonlinear optimal control problem
\begin{align}
&\min_{\vu_n, \vx_n} \left \{\Phi(\mathbf x_{N})+\sum_{n=0}^{N-1} L_n (\mathbf x_n,\mathbf u_n, n) \right \}
\label{eq:nonlinear_cost}\\
& \textrm{s.t.} \qquad \mathbf x_{n+1} - \mathbf f_n(\mathbf x_n,\mathbf u_n, n) = 0,\hspace{4mm} \mathbf x_0=\mathbf x_{init}
\label{eq:nonlinear_dynamics}
\end{align}
with state-vector $\mathbf{x}_n \in \mathbb R^m$ and control input vector $\mathbf{u}_n \in \mathbb R^p$. Let $L_n$ be the intermediate cost at time-step $n$ and $\Phi(\mathbf{x}_N)$ the terminal cost at the time horizon~$N$.

\subsection{Forming LQ subproblems}
\label{sec:formingLQsubproblems}
The optimal control law is computed in an iterative way. In each
iteration $k=0,1,2,\ldots$, we construct a LQ optimal control problem around the state trajectory 
$\vX^{[k]} = \{ \vx_0^{[k]}, \vx_1^{[k]}, \ldots, \vx_N^{[k]} \}$, with \mbox{$\vx_0^{[k]}=\mathbf{x}_{init}$} and the control trajectory 
$\vU^{[k]} = \{ \vu_0^{[k]}, \vu_1^{[k]}, \ldots, \vu_{N-1}^{[k]} \}$. 
In the first iteration, $k=0$, the LQ problem is hence constructed around the initial guesses for $\vX^{[0]}$, $\vU^{[0]}$. Possible initialization strategies are summarized in Section~\ref{sec:initialization}.

At each iteration, we numerically forward integrate all multiple shooting intervals using the respective control inputs $\vu_n^{[k]}$, starting at every state $\vx_n^{[k]} \ \forall n=0,1,\ldots, N-1$.
Fig.~\ref{fig:algorithms}c shows a sketch of the multiple-shooting intervals in GNMS, where the resulting state at the end of each interval is denoted $\vF^{[k]}(\vx_n^{[k]}, \vu_n^{[k]})$. Accordingly, we define the `defect' between the integrated trajectory segment and the next intermediate state $\vx_{n+1}^{[k]}$ as
\begin{equation}
 \vd_n^{[k]} = \vF^{[k]}(\vx_n^{[k]}, \vu_n^{[k]}) - \vx_{n+1}^{[k]} \ \text{.}
 \label{eq:defect}
\end{equation}
Defining state and control increments $\vdx_n^{[k]}$ and $\vdu_n^{[k]}$ for every single time-stage $n$, we can write the nonlinear system dynamics constraint~\eqref{eq:nonlinear_dynamics} in terms of the simulated interval as
\begin{equation}
\vx_{n+1}^{[k]} + \vdx_{n+1}^{[k]} - \vF^{[k]}(\vx_n^{[k]} + \vdx_n^{[k]}, \vu_n^{[k]} + \vdu_n^{[k]}) = 0
\label{eq:matching_condition}
\end{equation}
which can also be considered a matching condition which ensures the continuity of the state trajectory w.r.t. state and control increments. Performing a first-order Taylor expansion of Equation~\eqref{eq:matching_condition} w.r.t. $\vdx_n^{[k]}$ and $\vdu_n^{[k]}$, denoting the sensitivities w.r.t state and control $\vA_n$ and $\vB_n$ and taking into account the defects as defined by Equation~\eqref{eq:defect}, results in the following affine system dynamics constraint
\begin{equation}
\vdx_{n+1}^{[k]} - \vA_n^{[k]} \delta\vx_n^{[k]} - \vB_n^{[k]} \delta\vu_n^{[k]} - \vd_n^{[k]} = 0  \ \textrm{.} 
\label{eq:affine_constraint}
\end{equation}


Analogously performing a second-order Taylor expansion of the nonlinear cost function~\eqref{eq:nonlinear_cost} gives rise to the following LQ optimal control problem
\begin{align}
\min_{\vdu_n, \vdx_n} \bigg\{&q_N +\vdx_N^\top\vq_N+\tfrac{1}{2}\vdx_N^\top\vQ_N\vdx_N  \notag \\ 
	&+\sum_{n=0}^{N-1} q_n+ \vdx_n^\top\vq_n + \vdu_n^\top \vr_n+ \tfrac{1}{2}\vdx_n^\top\vQ_n\vdx_n \notag  \\
	& \quad +\tfrac{1}{2}\vdu_n^\top\vR_n\vdu_n + \vdu_n^\top\vP_n\vdx_n \bigg\} 
	\label{eq:linearized_cost}\\
\textrm{s.t.} \quad & \delta \vx_{n+1} = \vA_n \delta\vx_n + \vB_n\delta\vu_n +\vd_n \label{eq:affine_system} \\
&\mathbf x_0=\mathbf x_{init} \label{eq:initial_condition}
\end{align} 
%
where we assume $\vQ_n\textrm{,} \  \vQ_N \geq 0$ and $\vR_n > 0$. Here, and in the following subsection, we drop the superscript indices~$[k]$ for better readability.

\subsection{Computing the Optimal Control by Riccati Recursion}
\label{sec:riccati_recursion}
Considering the LQ subproblem~\eqref{eq:linearized_cost}-\eqref{eq:initial_condition}, the optimal control and state updates can be computed using a value-function approach. Assume a quadratic value function of the form
\begin{equation}
\label{eq:cost_to_go}
V_{n}(\delta \mathbf x_{n}) = s_{n} + {\delta \mathbf x_{n}^\top \mathbf s_{n}} 
 + \tfrac{1}{2} \delta \mathbf x_{n}^\top \mathbf S_{n} \delta \mathbf x_{n}
\end{equation}
with weighting matrices $\vS_n \in \mathbb R^{m \times m}$, $\vs_n \in \mathbb R^{m \times 1}$ and \mbox{$s_n \in \mathbb R$}. The optimal control update can be derived by minimizing the value function $V_n$ as a function of $\vdx_n$. 

As Equation~\eqref{eq:cost_to_go} is quadratic in $\vdx_{n+1}$ at time $n+1$, it remains quadratic during back-propagation in time, given the affine system dynamics and the linear-quadratic cost in Equations~\eqref{eq:linearized_cost}-\eqref{eq:initial_condition}. Due to Bellman's Principle of Optimality, the optimal control $\vdu_n^*$ at time $n$ can be computed from
\begin{align*}
&V^*_n(\vdx_n) = \min_{\vdu_n} \bigg[ q_n + \vdx^\top (\vq_n + \tfrac{1}{2} \vQ_n \vdx_n) 
+ \vdu_n^\top \vP_n \vdx_n  \notag \\
&+ \vdu_n^\top(\vr_n + \tfrac{1}{2} \vR_n \vdu_n)
+ V^*_{n+1}(\vA_n \vdx_n + \vB_n \vdu_n +\vd_n) \bigg]
\end{align*}
Inserting Equation~\eqref{eq:cost_to_go} and the affine system dynamics~\eqref{eq:affine_system} and minimizing the overall expression w.r.t. $\vdu_n$ leads to an optimal control update of the form
\begin{equation}
\vdu_n^*= -\vH_n^{-1} \vh_n - \vH_n^{-1} \vG_n \vdx_n
\label{eq:optimal_control_input}
\end{equation}
where we have defined
\begin{align*}
\vh_n & = \vr_n + \vB_n^\top (\vs_{n+1} + \vS_{n+1} \vd_n )\notag\\
\mathbf G_n & = \vP_n + \vB_n^\top \vS_{n+1} \vA_n \notag\\
\mathbf H_n & = \vR_n + \vB_n^\top \vS_{n+1} \vB_n
\end{align*}
and $\vB_n^\top \vS_{n+1} \vB_n + \vR_n > 0$.
After equating coefficients with a quadratic value function ansatz~\eqref{eq:cost_to_go} for $\vdx_n$, we define 
\mbox{$\vl_n = -\vH_n^{-1} \vh_n$}
and 
\mbox{$\vL_n = - \vH_n^{-1} \vG_n$}
and obtain the following recursive Riccati difference equations for $\vS_n$, $\vs_n$ and $s_n$
\begin{align}
\vS_n &= \vQ_n + \vA_n^\top \vS_{n+1} \vA_n - \vL_n^\top \vH_n \vL_n \label{eq:riccati_vS}\\
\vs_n &= \vq_n + \vA_n^\top (\vs_{n+1} + \vS_{n+1} \vd_n) \notag \\& \qquad + \vG_n^\top \vl_n + \vL_n^\top (\vh_n + \vH_n\vl_n) \label{eq:riccati_vs}\\
s_n &= q_n + s_{n+1} + \vd_n^\top \vs_{n+1} + \tfrac{1}{2} \vd_n^\top \vS_{n+1} \vd_n \notag \\
& \qquad + \vl_n^\top (\vh_n + \tfrac{1}{2} \vH_n \vl_n) \label{eq:riccati_s}
\end{align}
for $n \in {0,\ldots, N-1}$. For the final time-step~$N$ we obtain the terminal conditions \mbox{$\vS_N = \vQ_N$}, \mbox{$\vs_N = \vq_N$} and \mbox{$s_N = q_N$}, and the recursion is subsequently swept backwards. Note that Equation~\eqref{eq:riccati_s} does not contribute to the control update and can therefore be omitted in practice.

\subsection{Updating State and Control Trajectories}
Finally, using Equations~\eqref{eq:affine_system} and~\eqref{eq:optimal_control_input}, and readopting the superscript indices $[k]$ for the iteration count, we obtain equations for a forward sweep resulting in a full-step update for the control and state decision variables $\vX^{[k+1]}$, $\vU^{[k+1]}$
\begin{align}
\vu_n^{[k+1]} &= \vu_n^{[k]} + \vl_n^{[k]} + \vL_n^{[k]} (\vx_n^{[k+1]} - \vx_n^{[k]}) \label{eq:gnms_control_update}\\
\vx_{n+1}^{[k+1]} &= \vx_{n+1}^{[k]} + (\vA_n^{[k]} + \vB_n^{[k]} \vL_n^{[k]})(\vx_n^{[k+1]} - \vx_n^{[k]}) \notag \\  
& \quad + \vB_n^{[k]} \vl_n^{[k]} + \vd_n^{[k]} \label{eq:gnms_state_update}
\end{align}
with initial condition \mbox{$\vx_0^{[k+1]} = \vx_{init}$}. The updated decision variables are dynamically consistent w.r.t. the LQ subproblem dynamics. 
The nonlinear optimal control problem is solved iteratively, starting from Section~\ref{sec:formingLQsubproblems} and solving LQ subproblems at each iteration, until convergence.

\section{A Family of iLQR-GNMS Algorithms}
\label{sec:algorithms}
Equations~\eqref{eq:gnms_control_update} and~\eqref{eq:gnms_state_update} present the GNMS update rule where all states and controls (except for $\vx_{init}$) are decision variables. For every time-step, both states and controls are updated using a linear forward sweep.
Considering Equations~\eqref{eq:gnms_control_update} and~\eqref{eq:gnms_state_update}, we can now draw connections between GNMS and other existing algorithms and extend them to a bigger family of `hybrid' variants.
\subsection{Connection to iLQR and Single Shooting}
Interestingly, full-step iLQR employs the very same control update rule as in Equation~\eqref{eq:gnms_control_update}. In fact, GNMS can be transcribed into iLQR by substituting the state update equation~\eqref{eq:gnms_state_update} with a numeric forward integration of the nonlinear system~\eqref{eq:nonlinear_dynamics} using the time-varying state-feedback control law provided by Equation~\eqref{eq:gnms_control_update}.
In this case, the forward integration naturally results in a dynamically consistent state trajectory, all defects $\vd_n$ become zero and the formulation from section~\ref{sec:riccati_recursion} drops back to the well-known iLQR Riccati recursion.
Moreover, standard unconstrained single shooting can be recovered by additionally ignoring the state feedback gains and running the forward-integration purely open-loop.

\subsection{Hybrid Algorithms}
Consider a case where the overall time horizon $N$ is split into an integer number of multiple shooting intervals $M$ with length $l$ and \mbox{$1<M<N$}, while the control input discretization is kept at its original resolution.
Without loss of generality, let us assume that the MS integration intervals start at time indices $i \in \mathcal I$, with $\mathcal I = \{0, l, 2l,\ldots \}$. Fig.~\ref{fig:algorithms}d sketches an example of such a hybrid case with $l=3$. 
Every interval is simulated using the nonlinear system dynamics~\eqref{eq:nonlinear_dynamics} and the initial states and controls $\vx_i^{[k]}$ and $\vu_i^{[k]}$. All $\vx_j^{[k]}$ with $j \notin \mathcal I$ are \emph{overwritten} by the integration. For an open-loop forward integration, $\vU^{[k]}$ remains as is, but for a closed-loop forward integration, we additionally overwrite all $\vu_j^{[k]}$ with $j \notin \mathcal I$ using the given feedback control law. Note that in this case, the defect equation~\eqref{eq:defect} remains valid, but is zero along the multiple-shooting intervals. The only non-zero defects occur at $\vd_{i+l-1}^{[k]}, i \in \mathcal I$.
In this setting, the LQ approximation, Riccati recursion and state- and control updates~\eqref{eq:gnms_control_update}-\eqref{eq:gnms_state_update} can be performed as described before.
This gives rise to two `hybrid' GNMS variants:
\begin{itemize}
\item \emph{GNMS(M)}, using solely the feedforward control and thus performing an open-loop forward integration on each of the $M$ multiple shooting intervals, which themselves are multiples of the control interval. Herewith, standard single shooting is the limit case GNMS(1).
\item \emph{iLQR-GNMS(M)}, using the full state feedback controller~\eqref{eq:gnms_control_update} and a closed-loop forward integration of each multiple-shooting interval. In other words, this is equivalent to a multiple-shooting variant of iLQR. The standard iLQR algorithm is the limit case of iLQR-GNMS(1), with only one multiple-shooting interval.
\end{itemize}
Note that both GNMS($N$) and iLQR-GNMS($N$), with the number of multiple shooting-intervals being equal to the number of stages, revert to the standard GNMS formulation as introduced in Section~\ref{sec:gnms}. 
Table~\ref{tab:compTable} provides a compact overview of the algorithmic variants and compares their features.

\begin{table}
\caption{An overview of different GNMS-type methods.}
\label{tab:compTable}
\footnotesize
\tabcolsep=0.11cm
\begin{tabular}{ |m{7em} |c|c|c|c|c|  } 
\hline
            & SS    & iLQR  & GNMS  & GNMS($M$) & iLQR-GNMS($M$) \\ \hline \hline
closed-loop & --     & \checkmark     & --     & --         & \checkmark\\ \hline
No. intervals   & 1     & 1     & N     & M         & M\\ \hline
overwrite states by integration & \checkmark & \checkmark  & -- & \checkmark & \checkmark\\ \hline
need stable initial policy & \checkmark & \checkmark  & -- & depends & depends\\ \hline
\end{tabular}
\vspace{1ex} 

\raggedright (\checkmark = true, -- = false)
\end{table}
\subsection{Main Iteration and Implementation}
We emphasize that all algorithmic variants feature linear complexity in the time horizon, $O(N)$. All algorithms execute almost identical linear algebra operations during one major iteration and therefore have very similar computational effort.
Since the discussed family of GNMS algorithms only differs in a few features, it can be summarized in one framework, given in Algorithm~\ref{alg:general_algorithm}.
From a software-engineering perspective, the algorithmic variants are easy to implement and can all be treated at once, given a proper design of classes and interfaces. We provide an open-source C++ implementation of all discussed algorithms in~\cite{adrlCT}.

\subsection{Initialization}
\label{sec:initialization}
The GNMS variants listed in Table~\ref{tab:compTable} differ in their requirements for initialization. 
For iLQR and SS, the nominal state (and control) input trajectories are first updated through a forward integration.
This implies that for unstable systems, an initialization with a stabilizing initial control policy, which keeps the first rollout in the vicinity of the expected optimum, is essential.
For iLQR, the initially provided state trajectory $\vX^{[0]}$ serves as state reference trajectory for the feedback controller. For SS, it is irrelevant, except for the initial state. Common choices for SS and iLQR initial guesses are policies that stabilize the given initial state or draw the system towards the goal state, for example simple LQR or PD controllers. Generally, the increased efforts for initial guess design for SS and iLQR can be a significant disadvantage. In the worst case, a poor initial guess can lead to a local minimum with a solution far from desired behavior. 

Multiple-shooting algorithms, by contrast, offer greater flexibility and simplicity at initial guess design, and are often more robust w.r.t. bad initial policies. It is a well known fact that the convergence of multiple-shooting methods can be accelerated through an `educated' initial guess, such as direct interpolation between initial and desired final state. For the hybrid algorithms iLQR-GNMS($M$) and GNMS($M$) it often depends on the system characteristics if a stabilizing control policy is required, or if the multiple-shooting intervals are short enough to prevent significant divergence during integration.
In the video attachment~\cite{videoAttachment}, we show two simulation examples where initialization with a bad state-feedback controller significantly extends the runtime of iLQR compared to GNMS, or even causes iLQR to fail.

Note that, when all possible GNMS variants are initialized  with a dynamically consistent state trajectory and corresponding control trajectory, the defects for the first iteration are zero and the feedforward control updates are identical.

\begin{algorithm}[tpb] 
\caption{Generalized iLQR-GNMS($M$) Algorithm} 
\label{alg:general_algorithm}
\begin{algorithmic} 
\scriptsize 
\STATE \textbf{Given} 
\STATE - Nonlinear dynamics, cost function and initial state $\vx_{init}$ as given in~\eqref{eq:nonlinear_cost}-\eqref{eq:nonlinear_dynamics}
\STATE - Initial state trajectory $\vX^{[0]}$
\STATE - Initial feedforward trajectory $\vU^{[0]}$
\STATE - (Initial feedback law, if applicable)
\STATE - maximum total constraint violation $d_{max}$
\STATE - minimum relative cost change $J^{rel}_{min}$
\STATE \textbf{Prepare}
\STATE - set iteration count $k=0$
\STATE - split time horizon $N$ into $M$ MS integration intervals of length $l$, 
\STATE \hspace{1em}each starting at  an index $i \in \mathcal I$, $\mathcal I = \{0, l, 2l,\ldots \}$
\STATE - \textbf{Initial multiple-shooting rollouts}
\STATE \hspace{1em} - simulate $M$ intervals using the nonlinear system dynamics~\eqref{eq:nonlinear_dynamics} and initial
\STATE \hspace{2em} states and controls $\vx_i^{[0]}$ and $\vu_i^{[0]}$, overwrite all $\vx_j^{[0]}$ and $\vu_j^{[0]}$ for $j \notin \mathcal I$
\STATE \hspace{1em} - compute defects $\vd_n^{[0]}$ according to Equation~\eqref{eq:defect}.
\STATE \textbf{Repeat (main iteration)}
\STATE \hspace{1em} \textbf{LQ approximation}
\STATE \hspace{1em} - Linearize the dynamics along the trajectories, obtain the affine constraint~\eqref{eq:affine_constraint}
\STATE \hspace{1em} - Quadratize cost function along the trajectories to obtain~\eqref{eq:linearized_cost}
\STATE \hspace{1em} \textbf{Riccati backward sweep}
\STATE \hspace{1em} - Backwards solve the Riccati-like difference equations~\eqref{eq:riccati_vS}-\eqref{eq:riccati_vs} with 
\STATE \hspace{2em} boundary conditions $\vS_N = \vQ_N$ and $\vs_N = \vq_N$
\STATE \hspace{1em} \textbf{Linear forward sweep}
\STATE \hspace{1em} - compute state and control solution candidates $\vX^{[k+1]}$ and $\vU^{[k+1]}$ by 
\STATE \hspace{2em} forward sweeping Equations~\eqref{eq:gnms_control_update} and~\eqref{eq:gnms_state_update}.
\STATE \hspace{1em} \textbf{Rollout multiple-shooting intervals}
\STATE \hspace{1em} \textbf{IF}(\textit{open loop shooting})
\STATE \hspace{1em} \hspace{1em} - set feedback gain in Equation~\eqref{eq:gnms_control_update} to zero.
\STATE \hspace{1em} \textbf{ENDIF}
\STATE \hspace{1em} - simulate $M$ shooting intervals using nonlin. dynamics~\eqref{eq:nonlinear_dynamics}, controller~\eqref{eq:gnms_control_update}
\STATE \hspace{1em} \hspace{1em} and initial states and controls $\vx_i^{[k+1]}$ and $\vu_i^{[k+1]} \in \mathcal I$, 
\STATE \hspace{1em} \hspace{1em} overwrite all $\vx_j^{[k+1]}$ and $\vu_j^{[k+1]}$ for $j \notin \mathcal I$
\STATE \hspace{1em} - compute defects $\vd_n^{[k+1]}$ according to Equation~\eqref{eq:defect}
\STATE \hspace{1em} - compute cost $J^{[k+1]}$ by evaluating Equation~\eqref{eq:nonlinear_cost}
\STATE \hspace{1em} - increment iteration count $k$
\STATE \textbf{until} $|J^{[k]}-J^{[k-1]}|/J^{[k]} < J^{rel}_{min}$ \textbf{and} $\sum |\vd_n^{[k]}| < d_{max}$
\end{algorithmic} 
\end{algorithm}

\section{Results and Comparison}
\label{sec:simulation_results}
\subsection{An Illustrative, One-Dimensional System}
\label{sec:simpleSystem}
\begin{figure}
\centering
\includegraphics[width=0.999\columnwidth]{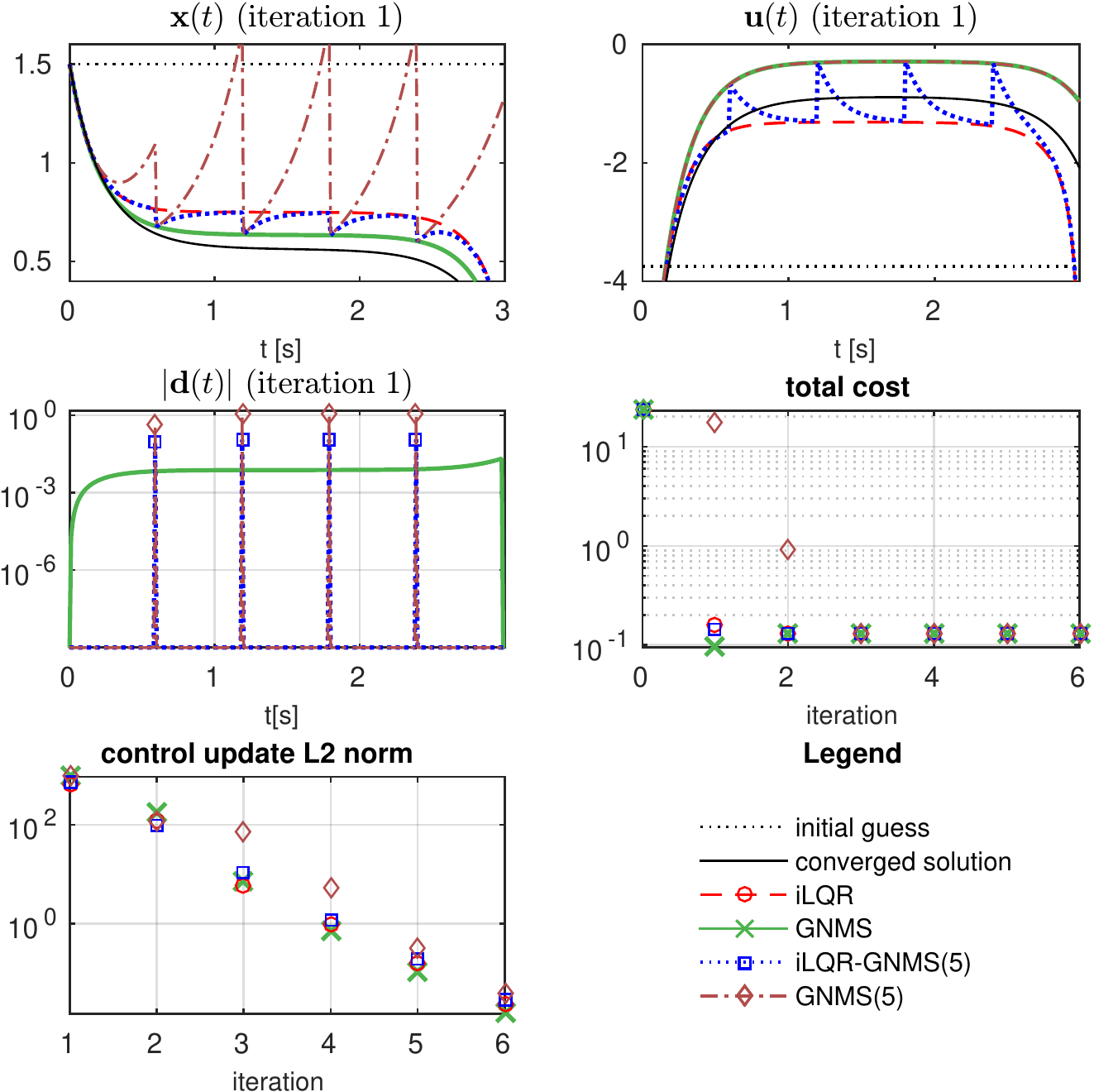}
\caption{
Results for a one-dimensional illustrative system, main text for detailed description.
}
\label{fig:simpleSystem}
\end{figure}
As an illustrative example, we present a simple one-dimensional system, which is slightly nonlinear, unstable and constructed to help the reader build an intuition about the methods. 
The system dynamics are
$\dot x = (1+x)x + u$, $x(0)=1.5$ 
and discretized with $\Delta t = 0.01$~s, $N=300$. The cost function is defined as quadratic cost of form~\eqref{eq:linearized_cost} with desired terminal state \mbox{$\vx_N^{des}= 0$}, $\vQ_N = 10$ and $\vR_n = 0.01$.
Fig.~\ref{fig:simpleSystem} shows results for iLQR, GNMS and the hybrid variants with five multiple-shooting intervals, GNMS(5) and iLQR-GNMS(5). We plot the state, control and defect trajectories for the first iteration of the algorithms, along with the initial guess and the converged solution. 

The state and control trajectories illustrate the relationship between the algorithms: the multiple-shooting intervals of GNMS(5) and iLQR-GNMS(5) start with states and controls lying on the respective GNMS trajectories. For GNMS(5), the system is simulated open-loop, the controls are identical to GNMS, and the state trajectories on the multiple-shooting intervals start to diverge. 
By contrast, for iLQR-GNMS(5), in every multiple-shooting interval both state and control trajectories converge asymptotically towards the simulated iLQR state and control trajectories. 
For the hybrid variants, a defect occurs every 0.6s, for GNMS, the defect is evenly distributed across all time intervals.
Due to the long shooting-intervals, GNMS(5) requires one iteration more to catch up with the other algorithms in terms of overall cost. Importantly, the control update plot shows that the asymptotic contraction rates, which are defined as
\begin{equation}
C=\lim_{k\rightarrow \infty} \nicefrac{|\vU^{[k+1]}-\vU^*|_2}{|\vU^{[k]}-\vU^*|_2}
\label{eq:asymptotic_contr_rate}
\end{equation}
are not the same. In this example, GNMS and GNMS(5) show better contraction than iLQR. Asymptotic contraction rates are investigated in more detail in Section~\ref{sec:contraction_rates}.

\subsection{Quadruped Trot Optimization Example}
\label{sec:quadrupedTO}

%

\begin{wrapfigure}{r}{0.49\columnwidth}
\includegraphics[width=0.49\columnwidth]{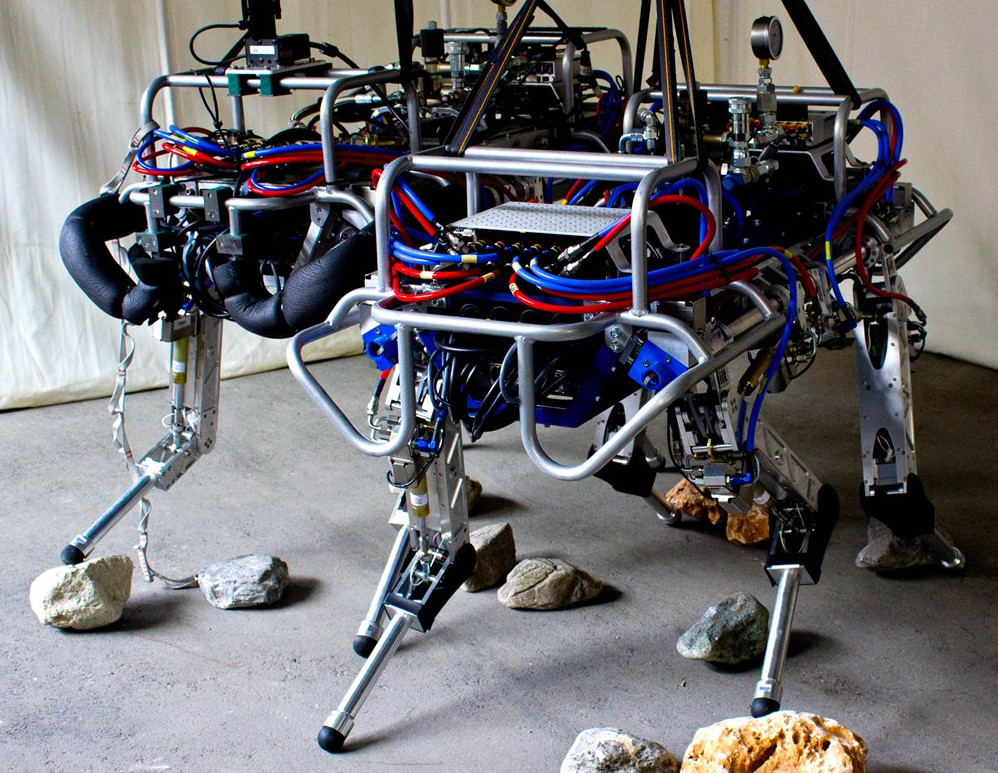}
\caption{The quadruped HyQ}
\label{fig:hyqs}
\vspace{-1em}
\end{wrapfigure}

The quadrupedal robot `HyQ'~\cite{semini:2011:hyqjournal} is an 18~DoF, floating-base underactuated robot subject to contacts with the environment, c.f. Fig.~\ref{fig:hyqs}. 
In this paper, the contacts are not
 incorporated as constraints, but added to the system dynamics using an explicit contact model. 
We employ a static, plain environment and a `soft' contact model, consisting of a nonlinear spring in surface-normal direction and a nonlinear damping term. The contact model is detailed in~\cite{Giftthaler2017autodiff}. Using this formulation, the contact force is a function of the current robot state only.
It is clear that such a soft contact model presents only a rough approximation of the complicated physics of contact, and also introduces a number of potential disadvantages such as increased stiffness and nonlinearity of the combined system dynamics.
However, the contact model allows a straight-forward computation of derivatives~\cite{Giftthaler2017autodiff}, which creates an ideal test-bed for comparing our shooting algorithms.
We obtain exact discrete sensitivities $\vA_n$ and $\vB_n$ through evaluating a sensitivity differential equation on the multiple-shooting intervals~\cite{dickinson1976sensitivity}.

The example task considered is the optimization of a periodic trotting gait. To achieve the trotting gait, we impose a time-varying quadratic penalty on the leg joint positions. 
Furthermore, we penalize the intermediate and final position of the robot's trunk and the intermediate and final velocities of the leg joints. For an in-depth description of the cost modelling to achieve different gait patterns the reader is referred to~\cite{neunert:2017:ral}.
In the following, the trotting gait optimization is used to compare the algorithms developed in this paper.
For a meaningful comparison, we initialize all algorithms with identical state trajectories and control policies. The initial guess corresponds to standing still in a steady state. We optimize over 36 states, 12 control inputs and a total time horizon of 2.5~seconds with $N=2500$.
\begin{figure}
\centering
\includegraphics[width=0.99\columnwidth]{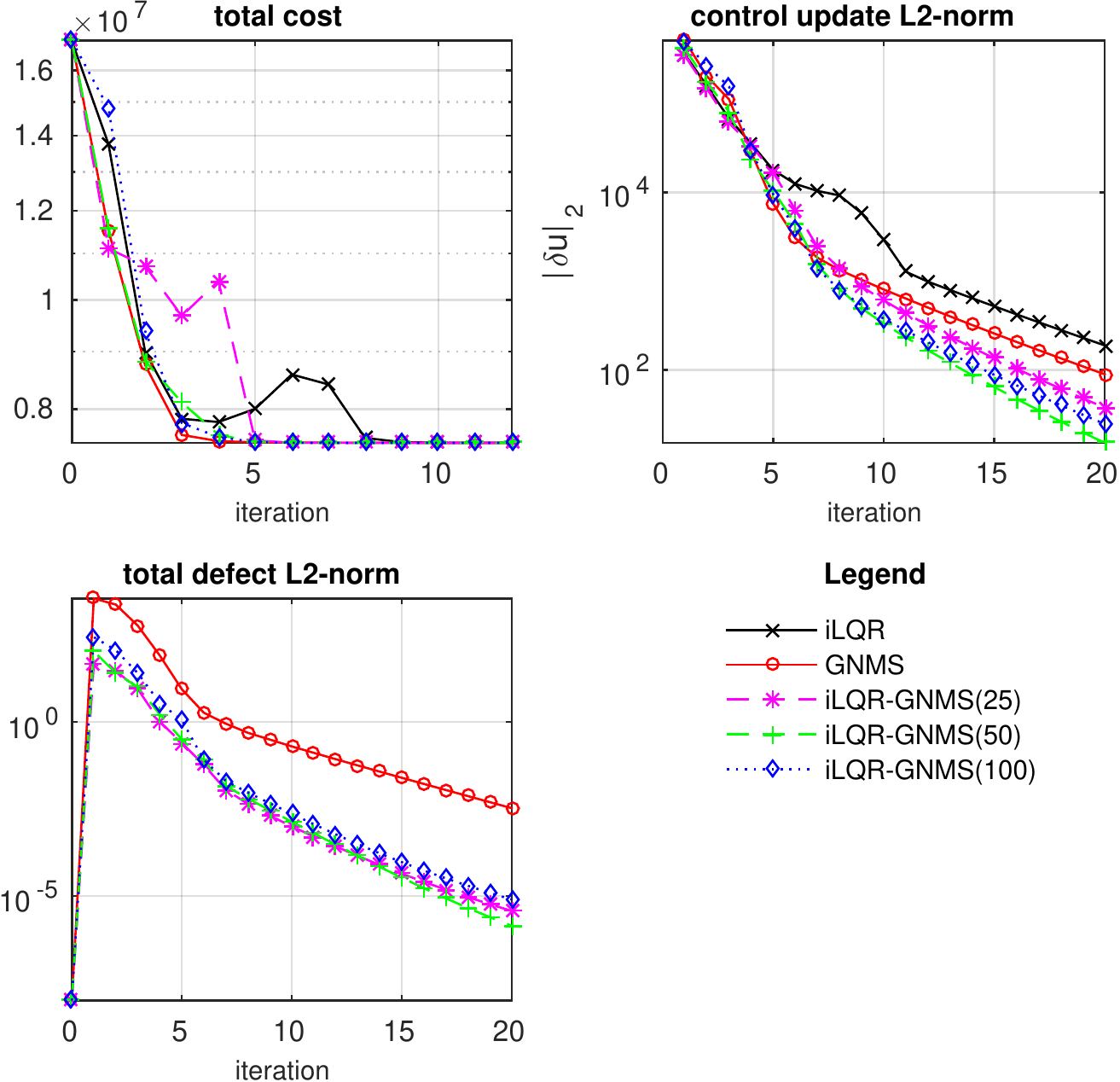}
\caption{Comparing different algorithms from the family of Gauss-Newton shooting methods w.r.t. cost descent, total defects and control update norm, illustrated with the example of a quadruped trot optimization problem.}
\label{fig:hyq_trot}
\end{figure}

Fig.~\ref{fig:hyq_trot} compares iLQR, GNMS, and iLQR-GNMS($M$) with three different numbers of multiple-shooting intervals in terms of cost descent, control update norms and total defects.
Note that SS and GNMS($M$) are unstable due to the strong instability of the system. All remaining algorithms converge to the same minimum within 20~iterations.
iLQR and iLQR-GNMS(25) show short phases of increasing cost, which we accept in this simulation example.
Since the provided initial guess is dynamically consistent, the initial defects are zero. GNMS, having the largest number of multiple-shooting intervals~(2500), also shows the largest total defect sum after the first iteration. The multiple-shooting iLQR methods, all having significantly fewer continuity constraints to enforce, feature a lower total defect.

As expected for a Gauss-Newton method, all approaches show linear convergence.
Considering the control update norms, we see that GNMS and iLQR feature a similar contraction rate for this example. In fact, the contraction rate of GNMS is slightly better, which is visually hard to distinguish here, but is detailed in following example.
For the hybrid multiple-shooting iLQR variants, we observe a significantly better contraction rate than for both iLQR and GNMS. When applying an identical termination criterion based on the relative change of the cost function and a defect threshold to all algorithms, all lifted methods converge in fewer iterations than iLQR. Furthermore, all displayed iLQR-GNMS($M$) variants converge notably faster than GNMS.
Screen recordings of the optimized trotting motions are provided in the video attachment~\cite{videoAttachment}.

\subsection{Local Contraction Rates for Quadruped Trot Tracking}
\label{sec:contraction_rates}
\begin{figure}
\centering
\includegraphics[width=0.99\columnwidth]{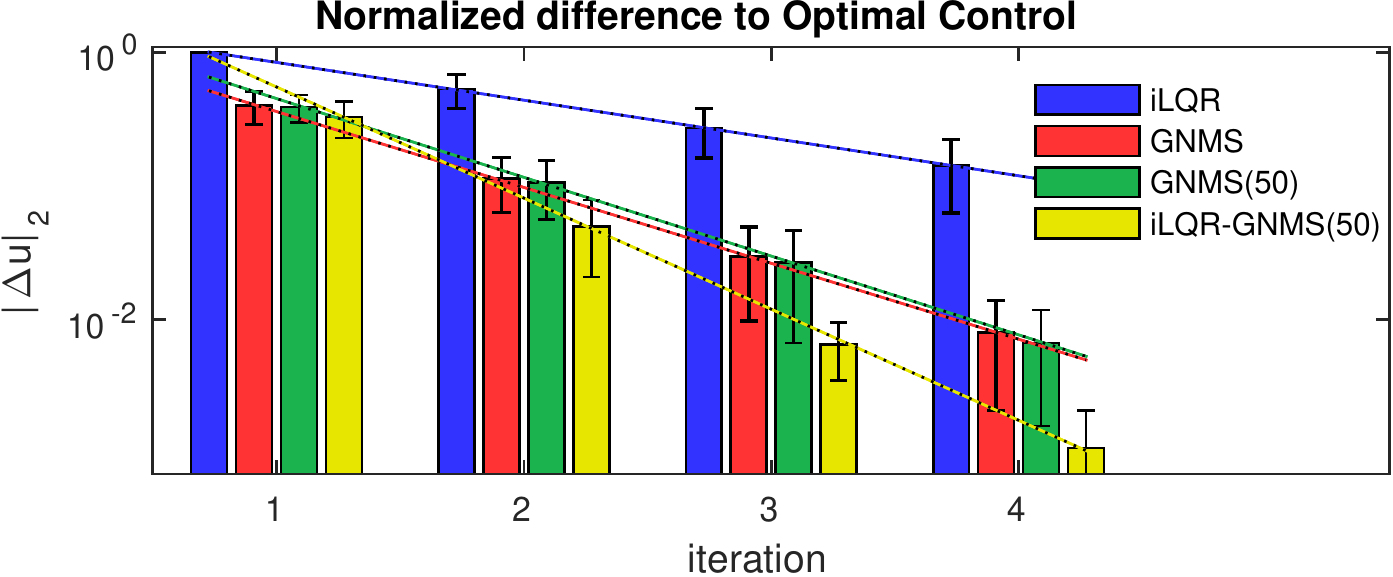}
\caption{
A statistical comparison of local contraction rates for four different algorithms. This plot shows the normalized difference between solutions to perturbed optimal control problems and a fully converged reference solution. The normalization is w.r.t. the iLQR control trajectory. The data is averaged from 1000 samples, the corresponding standard deviations are plotted as error-bars. Estimates for the contraction rates are indicated by straight lines.
GNMS outperforms iLQR in terms of local contraction rate, and gets closer to the true optimal solution in fewer iterations.
In contrast to the example in Section~\ref{sec:quadrupedTO}, GNMS(50) is stable and shows a contraction rate similar to GNMS. iLQR-GNMS(50) clearly outperforms all other algorithms with significantly better local contraction rate. 
After 4~iterations, it is on average 0.1\% away from the fully converged reference solution, while iLQR is on average 14\% away.}
\label{fig:hyq_contraction}
\end{figure}

While Section~\ref{sec:quadrupedTO} gives an optimization example for a single motion, starting with an initial guess far from the optimal solution, we now show a comparison based on statistical data from~$1000$ runs: the trotting gait from Section~\ref{sec:quadrupedTO} is now considered in a tracking MPC problem. All algorithms are initialized with an optimal, dynamically consistent solution, but the initial state is locally perturbed. The state perturbations are sampled from the hardware-experiments detailed in~\cite{neunert2017mpc}.
For every perturbation, we let different algorithms iterate until convergence.
Fig.~\ref{fig:hyq_contraction} compares average asymptotic contraction rates for four different algorithms. It shows the normalized difference between a fully converged optimal feedforward trajectory and trajectories obtained at previous iterations. Furthermore it shows first-order regressions approximating the local contraction rates, in terms of the slopes of the difference norms in the semi-logarithmic plot. 
It can be seen that GNMS outperforms iLQR in terms of local contraction rate.
GNMS(50) shows a contraction rate similar to GNMS. 
The example indicates better local convergence for iLQR-GNMS(50) than for classical iLQR, GNMS and GNMS(50). 

%

\begin{figure}
\centering
\includegraphics[width=0.99\columnwidth]{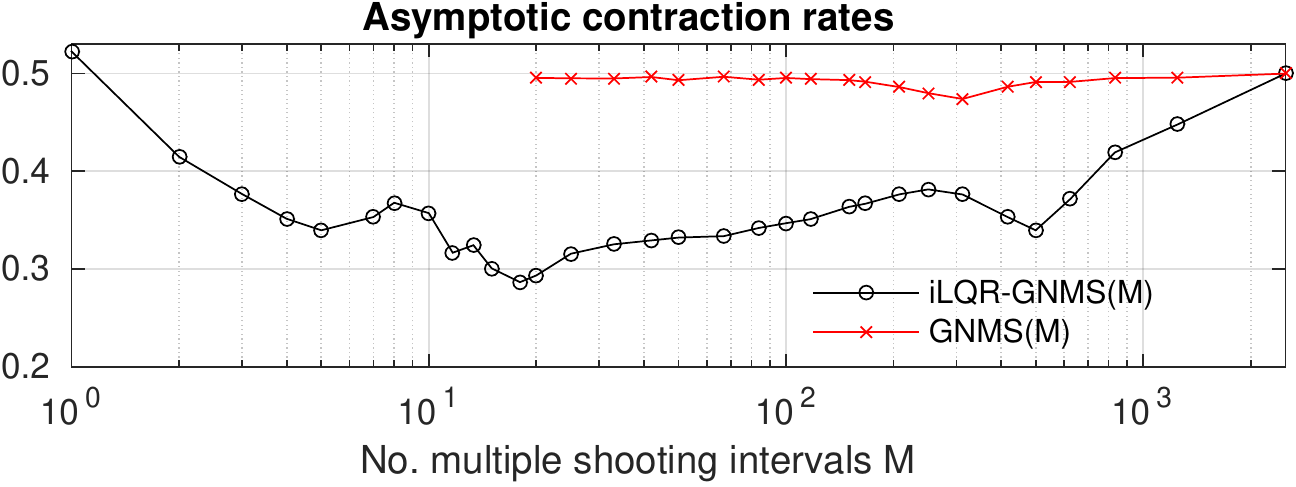}
\caption{
Comparing asymptotic contraction rates as a function of the number of multiple-shooting intervals~$M$. For GNMS($M$), the integration on the multiple-shooting intervals is only stable for $M\geq20$. For this range of `stable' multiple-shooting intervals, the contraction rates are almost constant and similar to the limiting case GNMS.
The iLQR multiple-shooting methods are stable for all $M$. Here, the contraction rates for intermediate numbers of multiple-shooting intervals are distinctively better than for the limit cases iLQR and GNMS.
}
\label{fig:asymptotic_contraction_rates}
\end{figure}

Fig.~\ref{fig:asymptotic_contraction_rates} generalizes the result from Fig.~\ref{fig:hyq_contraction} for a range of multiple-shooting intervals~$M$, showing numerically approximated asymptotic contraction rates, Equation~\eqref{eq:asymptotic_contr_rate}, as a function of~$M$. Again, GNMS($M$) is unstable for overly long multiple-shooting intervals, similar to the limiting case open-loop single shooting.
For closed-loop shooting, the asymptotic contraction rates for all multiple-shooting variants are better than for iLQR, and the contraction rates for the hybrid variants outperform the limiting case GNMS. In this example, the relative improvement over iLQR is up to a factor two.
Note that the resulting iLQR-GNMS($M$) contraction rates differ slightly from the ones in Fig.~\ref{fig:hyq_trot} for $M=25, 50, 100$, which is due to the different problem setting. However, both experiments exhibit the same trend.

\subsection{Nonlinear MPC on HyQ}
The suitability of iLQR for nonlinear model-predictive control (NMPC) in robotics applications has been shown many times before, in~\cite{koenemann2015whole,neunert16hexrotor,giftthaler2017efficient}. 
In this section, we show that GNMS and its hybrid variants are even more promising for NMPC applications. First, they converge faster to the optimal solution, c.f. Section~\ref{sec:contraction_rates}. A second advantage of the multiple-shooting variants of the presented algorithms is that the forward integrations can be parallelized. Therefore, the achievable MPC cycle time decreases approximately linearly with the number of available CPU cores. By combining faster update rates with better contraction rate, our multiple-shooting algorithms outperform classical iLQR-NMPC. 

In the following simulation example, we compare the NMPC performance of our Gauss-Newton shooting algorithms against iLQR-NMPC in a HyQ simulation environment. In each NMPC cycle, we run an adapted version of the main iteration in Algorithm~\ref{alg:general_algorithm} and `warm-start' it with the previous solution. In such a setting, we can separate an NMPC iteration into a `preparation' and a `feedback' phase~\cite{DIEHL2002}, thus minimizing the latency between receiving a state-measurement and sending an updated policy to the control system. Our NMPC loop is described in Algorithm~\ref{alg:gnms_mpc_algorithm}.
\begin{algorithm}[tpb] 
\caption{iLQR-GNMS($M$)-NMPC Algorithm} 
\label{alg:gnms_mpc_algorithm}
\begin{algorithmic} 
\scriptsize 
\STATE \textbf{Given}
\STATE - cost function~\eqref{eq:nonlinear_cost} and system dynamics~\eqref{eq:nonlinear_dynamics}.
\STATE - receding MPC time horizon $N$
\STATE - number of multiple-shooting intervals $M$ with length $l$
\STATE - initial state and control trajectories \mbox{$\vX = \{ \vx_0, \vx_1, \ldots, \vx_N\}$}
\STATE \hspace{1em}\mbox{$\vU = \{ \vu_0, \vu_1, \ldots, \vu_{N-1}\}$}, state-feedback controller $\vu_n(\vx)$ of form~\eqref{eq:optimal_control_input}
\STATE \textbf{Repeat Online:}
\STATE \textit{Feedback phase}
\STATE - get state measurement $\vx_{meas}$.
\STATE - forward integrate system dynamics~\eqref{eq:nonlinear_dynamics} with \mbox{$\vx_0 = \vx_{meas}$} on the first 
\STATE \hspace{1em}multiple-shooting interval, compute $\vA_{0,\ldots,l-1}, \ \vB_{0,\ldots,l-1}$, defect $\vd_{l-1}$
\STATE - quadratize cost function~\eqref{eq:nonlinear_cost} around $\vX$ and $\vU$ for control stages $1,\ldots, l-1$.
\STATE - solve LQ optimal control problem using a Riccati backward sweep
\STATE - retrieve updated control policy $\vu_n^{+}(\vx)$ and updated trajectories $\vU^+$, $\vX^+$.
\STATE - send policy $\vu_n^{+}(\vx)$ and $\vX^+$ to the control system
\STATE \textit{Preparation phase}
\STATE - update: $\vu_n(\vx) \leftarrow \vu_n^{+}(\vx)$, $\vX \leftarrow \vX^+$, $\vU \leftarrow \vU^+$
\STATE - forward integrate system dynamics~\eqref{eq:nonlinear_dynamics} for the multiple-shooting
\STATE \hspace{1em}intervals 1 to $M$, obtain sensitivities $\vA_l,\ldots \vA_{N-1}$, $\vB_l,\ldots, \vB_{N-1}$
\STATE \hspace{1em} and defects $\vd_{l,\ldots, N-1}$
\STATE - quadratize cost function~\eqref{eq:nonlinear_dynamics} around $\vX$, $\vU$ for multiple-shooting intervals $l$ to~$N$.
\end{algorithmic} 
\end{algorithm}

\begin{figure}
\centering
\includegraphics[width=0.99\columnwidth]{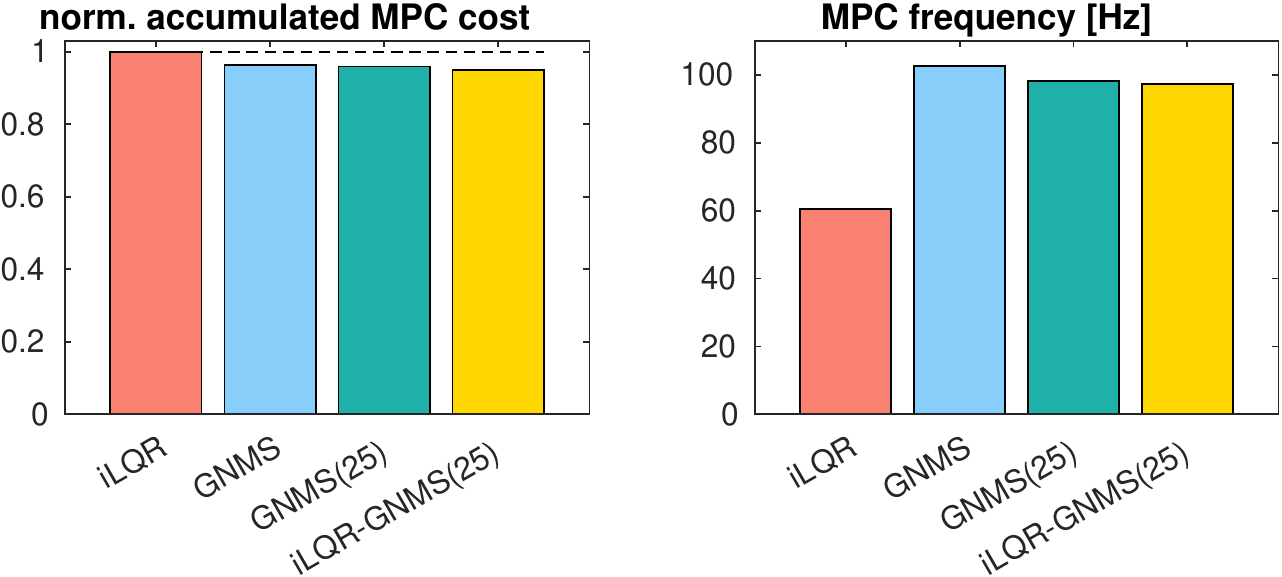}
\caption{Comparing four different solvers for HyQ-MPC. The left plot compares costs accumulated over 18.0 seconds of trotting-MPC execution. iLQR shows the highest accumulated cost. The multiple-shooting variants outperform iLQR, but the relative cost difference is small. GNMS runs at 96.3\% of the iLQR cost, GNMS(25) at 95.9\% and iLQR-GNMS(25) at 95.0\%.
Regarding the achieved average MPC frequencies, shown on the right, we find up to 40\% higher frequencies for the multiple-shooting variants. The maximum average MPC frequency, 103 Hz, is achieved using GNMS.}
\label{fig:hyq_mpc}
\end{figure}

In this experiment, we run a trotting gait on HyQ, in closed-loop MPC in a simulation environment~\cite{Schaal__2009}. For the NMPC optimal control problem, we choose a time-step size of 4~ms and \mbox{$N=125$}. We parallelize the integration of all multiple-shooting intervals and the sensitivity computation on four cores, and run both simulator and MPC controller on the same notebook equipped with an Intel Core~i7 (2.8~GHz) processor. 
For four different algorithmic combinations, we record the executed trot under identical conditions for 18~seconds and compute the resulting accumulated intermediate cost. A summary of the achieved average cost and NMPC frequencies is given in Fig.~\ref{fig:hyq_mpc}. In these experiments, iLQR results in the highest accumulated cost. The multiple-shooting variants outperform iLQR, with relative cost differences up to 5\%. At the same time, due to shorter runtimes, the multiple-shooting variants achieve up to 40\% higher MPC frequencies, with a maximum of 103~Hz.

In our simulation, all four algorithm variants run in a stable and robust fashion. The relatively small cost difference is an indicator of better convergence, but the main reason why the multiple-shooting variants should be preferred over iLQR in real-world applications, is the superior control bandwidth.
The algorithms in this paper have been validated in hardware experiments on two different quadruped platforms, where a variety of motions and gaits was implemented. However, a in-depth description of the experimental setups, the optimized computational framework and practical tuning considerations are beyond the scope of this paper. The interested reader is therefore referred to~\cite{neunert2017mpc}, where we apply the GNMS-algorithm for full-body NMPC on the quadrupeds HyQ and ANYmal~\cite{hutter2016anymal}, explain the robotic setup in detail and present a variety of hardware experiments. As outlook, a video sequence of GNMS-NMPC running on hardware is provided in the attachment~\cite{videoAttachment}.

\section{Conclusion and Outlook}
\label{sec:Discussion}
In this paper, we have shown how the well-known iLQR algorithm can be lifted and transformed into Gauss-Newton Multiple Shooting, GNMS. We have generalized the concept to form a family of Gauss-Newton shooting algorithms, which can be distinguished into sequential and simultaneous algorithms and closed and open-loop algorithms. Some algorithms partially overwrite decision variables by means of a numeric forward integration. All presented variants have approximately the same computational cost and feature linear time-complexity. Furthermore, all discussed algorithms share a large number of computational routines, and it is not difficult to implement all of the presented variants in a single software framework.

We have compared the performance of the algorithms in different simulation experiments, which indicate that the lifted algorithms can outperform classical iLQR.
While not included in this paper for reasons of compactness, similar results were obtained for other rigid-body dynamic systems including a 6~DoF fixed-base arm model. 
A more fundamental investigation for formalizing the conditions that result in improved convergence rates for GNMS($M$) and iLQR-GNMS($M$) is subject to ongoing work.

In the application examples, we limited the comparison to full-step variants of all considered algorithms. However, for even more nonlinear dynamics or cost functions, where the LQ optimal control problem is a bad approximation to the nonlinear problem, a globalization strategy may be required. For single-shooting methods, a straight-forward solution is to employ a line-search scheme. This is simple to implement, as it is sufficient to search over the cost for different control update step-sizes. For multiple-shooting approaches, however, there are additional continuity constraints, and we need to line-search over a merit-function which trades off the costs and defects. It is typically required to introduce additional tuning variables or to implement a complex filter-scheme~\cite{nocedal}.
Our open-source reference implementation~\cite{adrlCT} provides a line-search scheme using a simple merit function.

For complex robot trajectory optimization problems, we do not recommend to generally prioritize one of the presented algorithms over another. While the multiple-shooting algorithms allow for advanced initialization strategies and are more robust w.r.t. bad initial guesses, they may require slightly more tuning efforts when the full-step algorithm is not sufficient.
By contrast, in NMPC applications with well-defined cost functions and using warm-starting, additional globalization steps are rarely required at all. Here, our multiple-shooting algorithms offer significant advantages, better local contraction rates and much shorter runtimes.

The focus of this paper is on unconstrained optimal control problems without general (in)equality path constraints. It is obvious that the lifting approach naturally transfers to equality-constrained variants of iLQR, such as~\cite{sideris2011riccati}.
The inclusion of general (in)equality path constraints is part of ongoing work. One option to include them in the existing framework is to replace the standard Riccati backward sweep with a dedicated solver for constrained LQ optimal control problems~\cite{frisonalgorithms}. In this way, general (in)equality path constraints can be included while keeping linear time-complexity.

While this work treats algorithms using a Gauss-Newton Hessian approximation, it similarly transfers to exact-Hessian approaches, resulting in a multiple-shooting DDP algorithm combining the advantages of simultaneous methods, quadratic convergence and closed-loop integration.

\section*{Acknowledgements}
\small
The authors would like to thank Dimitris Kouzoupis, Gianluca Frison, Mario Zanon and Timothy Sandy for fruitful discussions. 
This research was supported by the Swiss National Science Foundation through the NCCR Digital Fabrication, the NCCR Robotics and a Professorship Award to Jonas Buchli.
Further, this research was supported by the EU via FP7-ITN-TEMPO (607 957) and H2020-ITN-AWESCO (642 682), by the Federal Ministry for Economic Affairs and Energy (BMWi) via eco4wind and DyConPV, and by DFG via Research Unit FOR 2401.

\bibliographystyle{ieeetr}
\bibliography{refs}

\end{document}

%% file: mathdef.tex
\newcommand{\vd}{\mathbf d}

\newcommand{\vh}{\mathbf h}

\newcommand{\vl}{\mathbf l}

\newcommand{\vq}{\mathbf q}
\newcommand{\vr}{\mathbf r}
\newcommand{\vs}{\mathbf s}

\newcommand{\vu}{\mathbf u}

\newcommand{\vx}{\mathbf x}

\newcommand{\vdx}{\delta \mathbf x}
\newcommand{\vdu}{\delta \mathbf u}

\newcommand{\vA}{\mathbf A}
\newcommand{\vB}{\mathbf B}

\newcommand{\vF}{\mathbf F}
\newcommand{\vG}{\mathbf G}
\newcommand{\vH}{\mathbf H}

\newcommand{\vL}{\mathbf L}

\newcommand{\vP}{\mathbf P}
\newcommand{\vQ}{\mathbf Q}
\newcommand{\vR}{\mathbf R}
\newcommand{\vS}{\mathbf S}

\newcommand{\vU}{\mathbf U}

\newcommand{\vX}{\mathbf X}

